# Observation of Magnetostatic Surface Spin Wave Solitons in Yttrium Iron Garnet Thin Film

Simin Pang,[1,#] Zhengyi Li,[2,4,#] Ziyu Wang,[5,#] Yanpei Lv,[1,3] Feilong Song,[1] Peng Yan,[4,*] and Jun Zhang[1,3,†]

[1] *State Key Laboratory of Semiconductor Physics and Chip Technologies, Institute of Semiconductors, Chinese Academy of Sciences, Beijing 100083, China*
[2] *Jiangsu Key Laboratory of Opto-Electronic Technology, School of Physics and Technology, Nanjing Normal University, Nanjing 210046, China*
[3] *Center of Materials Science and Optoelectronics Engineering, University of Chinese Academy of Sciences, Beijing 100049, China*
[4] *School of Physics and State Key Laboratory of Electronic Thin Films and Integrated Devices, University of Electronic Science and Technology of China, Chengdu 610054, China*
[5] *School of Physics and Laboratory of Zhongyuan Light, Zhengzhou University, Zhengzhou 450001, China*

[#] These authors contributed equally.

Magnetostatic surface spin wave (MSSW) solitons hold great promise for magnonic information processing, but their existence has long been debated. In this work, we resolve this issue by advanced time-resolved Brillouin light scattering (TR-BLS) spectroscopy. We observe long-period MSSW soliton trains in yttrium iron garnet (YIG) thin films by demonstrating their quasiparticle behavior, mode beating, and periodic modulations. We reveal that the MSSW soliton originates from the dipole gap mechanism with a unique comb-like frequency spectrum caused by the four-magnon process. By varying the microwave power, we find significant changes in soliton periods linked to the spin wave dispersion renormalization. Additionally, we report exotic transverse solitons across the YIG thickness. These findings deepen the understanding of nonlinear physics, and pave the way for spin-wave-soliton-based information technologies.

Spin waves show significant potential for future information processing and spintronic/magnonic applications due to their ability to propagate without Joule heating, enabling ultra-low power consumption [1-6]. However, their susceptibility to damping and distortion in dispersive media limits their long-distance transport. One promising solution is the spin-wave soliton—a localized wave packet that can propagate without losing its shape, amplitude, and speed through a balance between nonlinearity and dispersion [7]. Spin-wave solitons manifest in various forms, including envelope solitons [8-12], 'Möbius' solitons [13], bullets [14-16], magnetic droplets [17-19], domain walls [20, 21], vortices [22, 23], skyrmions [24-26], and hopfions [27-29]. Their exceptional stability and highly tunable dynamics make them promising candidates for a wide range of applications, including radio-frequency device [30, 31], logic [32, 33], racetrack memories [34, 35], spintronic unidirectional vertical shift register [36], and neuromorphic computing [37-39].

Among various spin-wave modes, magnetostatic surface spin waves (MSSWs) are particularly attractive due to their non-reciprocal nature [40, 41]. This unique characteristic makes MSSW solitons highly relevant for non-reciprocal communication systems, magnetic storage, and low-power logic devices. However, previous theory suggests that MSSW soliton is unlikely to happen, because the product of the dispersion ($D = \frac{\partial^2 \omega}{\partial k^2}$) and nonlinearity ($N = \frac{\partial \omega}{\partial |\psi|^2}$) is positive, according to the Lighthill criterion, with $\omega$, $k$, and $\psi$ being the angular frequency, wavevector, and amplitude of the spin wave, respectively [42, 43]. However, few recent experiments claimed evidence of soliton-like excitations for MSSWs [44, 45]. This thus brings about an outstanding issue of whether the MSSW soliton exists or not. Resolving this issue is critical, as it not only addresses the aforementioned fundamental question but also provides insights into the real-world application of non-reciprocal spin waves. Advanced experimental techniques are essential. Brillouin light scattering (BLS) spectroscopy offers a powerful approach to tackle this puzzle. Unlike conventional microwave transmission methods, BLS spectroscopy can detect both Stokes and anti-Stokes components, enabling the study of quasiparticle characteristics and nonreciprocity between them, along with the nonlinear dynamics in both time and space domains [46-48]. By capturing nonlinear interactions, such as three- and four-magnon scattering processes, BLS spectroscopy provides a unique feature to identify the underlying mechanisms for classification purposes.

In this Letter, we present conclusive evidence for the existence of MSSW solitons in yttrium iron garnet (YIG) thin films by time-resolved Brillouin light scattering (TR-BLS) spectroscopy. We reveal that dipole gaps in the spin wave dispersion repair the Lighthill criterion, which enables the generation of MSSW solitons. We observed multiple periodicities in the soliton wave, induced by four-magnon scatterings that modulate both the soliton's temporal behavior and its spectral characteristics. We show that the soliton periods vary by adjusting the microwave power due to the spin wave dispersion renormalization. Notably, we observe the emergence of transverse solitons across the YIG stripe, which display spatial periodicity—a new phenomenon that adds complexity to our understanding of MSSW dynamics. Micromagnetic simulations compare well with experimental results. Our findings provide a firm answer to the existence of MSSW solitons, which significantly deepen our understanding of the nonlinearity in non-reciprocal spin-wave dynamics, and pave the way for the development of advanced magnonic devices and wave-based computing.

*Contact author: yan@uestc.edu.cn

†Contact author: zhangjwill@semi.ac.cn

Figure 1(a) shows the schematic diagram of the experimental setup for spin-wave soliton excitation in YIG film and detection using space- and time-resolved BLS spectroscopy. The studies were performed on an optically transparent YIG film with a thickness of 4.2 μm, width of 2 mm, and length of 20 mm. A microstrip resonator placed beneath the YIG film excites spin waves by applying a microwave current with pumping frequency $f_p$ and power $P$. A laser beam is focused on the resonator, and the scattered light is collected and directed to the interferometer for analysis. A uniform static magnetic field with a nominal value of 460 Oe was applied in-plane and perpendicular to the length of the YIG strip ($y$-axis), saturating the magnetization, which supports the propagation of MSSWs. We proposed that by manipulating the sign of $D$, MSSW solitons can be successfully excited. The spin wave dispersion calculated for the YIG strip, shown in Fig. 1(b), illustrates the coupling between MSSWs and perpendicular standing spin waves, which results in a series of dipole gaps. These dipole gaps lead to dramatic changes in the dispersion parameters $D$, where the upper band gap corresponds to positive $D$ and the lower band gap to negative $D$, as shown in Fig. 1(c). By choosing the frequency within the upper band gap, we are able to excite MSSW solitons, with the nonlinear parameter $N$ remaining negative [49, 50]. The details of micromagnetic simulations are provided in Supplemental Material [51].

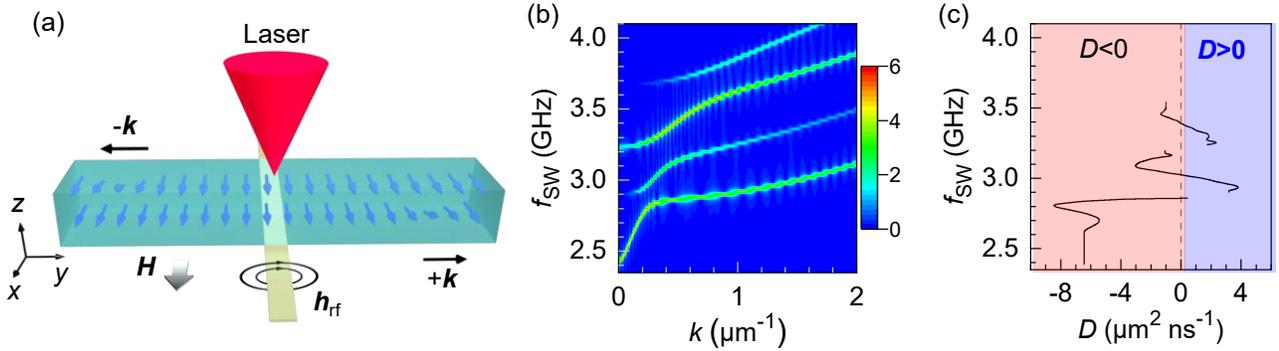

FIG. 1. (a) The experimental set-up for time- and space-resolved BLS measurements with microwave pumping. (b) Color plot of simulated MSSW dispersion relation with dipole gaps for the YIG stripe. (c) Extracted dispersion parameters $D$ from (b). The red (blue) transparent background indicates that $D < 0$ ($D > 0$). When $D > 0$, the Lighthill condition is satisfied with the nonlinear parameter $N$ remaining negative.

In the experiments, we selected the excitations with $f_p = 3.6$ GHz, where the Lighthill criterion is recovered. To investigate the temporal evolution of the excited spin waves, we performed the TR-BLS measurements with different microwave pulse widths ($w$), as shown in Figs. 2(a)-2(c). The measurements were done with a microwave power of 10 W, which refers to the nominal output power from the microwave source. We estimated that the power delivered to the sample is approximately 45 mW. The details of the power estimation are provided in Supplemental Material [51]. The signals were integrated over a frequency range from 2.5 to 4.75 GHz (Stokes side), around $f_p$. As shown in Fig. 2(a), the signal appears at approximately 115 ns and disappears around 615 ns for a pulse width of 0.5 μs. The rising and falling edges correspond to the switching on and off of the microwave excitation at the microstrip, respectively. In the upper panel of Fig. 2(a), the TR-BLS intensity displays periodic patterns with periods $t_1 \approx 87$ ns and $t_2 \approx 175$ ns, marked by two double arrows. Zooming in on the spectrum (dotted red square), as seen in the lower panel of Fig. 2(a), the signal persists even after the microwave source is switched off, revealing a shorter period $t_3 \approx 10$ ns. To explore the behavior over a longer time scale, we set $w$ to 10 μs and the pulse period ($T$) to 20 μs, maintaining the cycle duty. More obvious interference patterns are observed, forming the spin wave envelope soliton [upper panel of Fig. 2(b)]. In the zoom-in spectra shown in the lower panel of Fig. 2(b), a secondary modulation with a significantly longer period ($t_4 \approx 1149$ ns), marked by two dashed-dotted lines, appears on top of the first modulation ($t_2$). The corresponding fast Fourier transform (FFT) spectrum, displayed in the upper panel of Fig. 2(d), identifies three sets of duration time (modulation frequency $\Delta f$): $t_4 \approx 1149$ ns (0.87 MHz), $t_2 \approx 175$ ns (5.72 MHz), $t_{2-} \approx 206$ ns (4.85 MHz), $t_{2+} \approx 152$ ns (6.59 MHz), $t_1 \approx 87$ ns (11.44 MHz), $t_{1-} \approx 95$ ns (10.57 MHz), and $t_{1+} \approx 81$ ns (12.31 MHz). The corresponding modulation frequencies are marked with balls in the lower panel of Fig. 2(d), exhibiting the comb-like pattern with a frequency spacing of 0.87 MHz ($\Delta f_4$), marked by the double arrows.

Self-modulational instability (SMI) is responsible for the formation of comb structure here [53], with the underlying physical process involving four-magnon scattering. As shown in Fig. 2(e), the origin of $t_4$ arises from the scattering of two initial pumping magnons characterized by $f_p$ and $k_p$ (wavevector of the pumping magnons), into a pair of secondary magnons $f_4^+$ ($k_4^+$) and $f_4^-$ ($k_4^-$), following the conservation of energy and momentum. The $f_4^-$ mode corresponds to the bottom of the $n_1$ branch, where dipole gaps are present. The frequency difference between $f_p$ and $f_4^-$ ($f_4^+$) is 0.87 MHz ($\Delta f_4$), which matches the reciprocal of $t_4$. For $t_2 \approx 175$ ns, the corresponding modulation frequency is 5.72 MHz ($\Delta f_2 = f_p - f_2^-$). Here, $f_2^-$ ($f_2^+$) represents the mode on the $n_1 - 1$ ($n_1 + 1$) branch, with the same wavevector as the initial mode. In these branches, four-magnon scattering occurs via the process $2f_p = f_2^+ + f_2^-$. Specifically, our simulations reproduce such four-magnon scattering processes responsible for the soliton period $t_2$ and $t_4$, as shown in Supplemental Material [51]. Additionally, interference between $f_2^-$ and $f_4^-$ produces a new period of $t_{2-} \approx 206$ ns (4.85 MHz), while beating between $f_2^+$ and $f_4^-$ leads to $t_{2+} \approx 152$ ns (6.59 MHz). When the initial mode is sufficiently strong, it can further interact with modes on the $n_1 - 2$ and $n_1 + 2$ branches, which share the same wavevector with $f_p$, via the four-wave process: $2f_p = f_1^+ + f_1^-$. The frequency difference between $f_p$ and $f_1^-$ ($f_1^+$) is 11.44 MHz ($\Delta f_1$), corresponding to the observed $t_1 \approx 87$ ns. Similarly, interference between $f_1^-$ ($f_1^+$) and $f_4^-$ generates periods of $t_{1-}$ ($t_{1+}$), which contributes to the overall comb-like spectrum. Finally, $t_3 \approx 10$ ns ($\Delta f_3 \approx 97.35$ MHz) corresponds to the frequency difference between $f_p$ and the mode at the $n_2$ branch ($f_3^-$), satisfying the relation $2f_p = f_3^+ + f_3^-$ and $2k_p = k_3^+ + k_3^-$. When the pulse width $w$ is further increased to 30 μs and 100 μs, a soliton train sustains up to 50 μs [Fig. 2(c)].

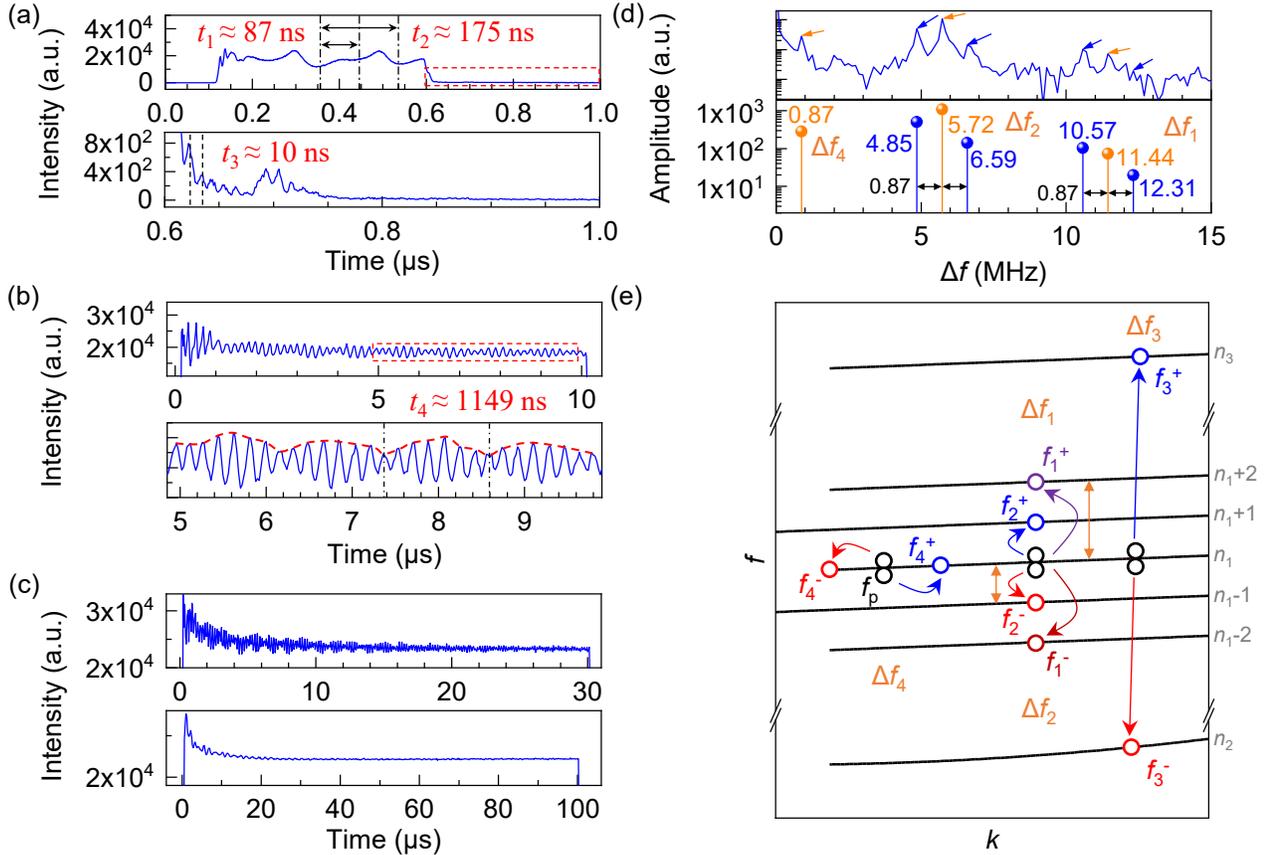

FIG. 2. Microwave pulse width-dependent TR-BLS spectra with a pulse width $w$ (pulse period $T$) of (a) 0.5 μs (1 μs), (b) 10 μs (20 μs), and (c) 30 μs (60 μs) [upper panel] and 100 μs (200 μs) [lower panel]. The lower panels in (a) and (b) correspond to the spectra indicated by the red dotted square in the upper panels. Measurements were performed at a microwave pumping frequency $f_p = 3.6$ GHz and power $P = 10$ W (nominal output power from the microwave source). (d) FFT result of (b), with the modulation frequency denoted by the blue balls in the lower panel. (e) Four-magnon scattering processes responsible for the MSSW soliton periods.

To investigate the power threshold for soliton formation and the variation of soliton periods, we conducted microwave power-dependent TR-BLS spectra measurements, as shown in Figs. 3(a)-3(e). At low microwave power (nominal output power from the microwave source $P = 0.8$ W), an approximately constant time-domain signal was observed. As $P$ increases to 1 W, the steepening effect due to the nonlinearity counterbalances the dispersive spreading [8, 9, 44], and long-time soliton trains begin to form [Fig. 3(b)]. The small period $t_2$ is the most dominant one in the time domain. Additionally, a previously unobserved period $t_5 \approx 85$ ns (11.75 MHz) appears, suggesting the involvement of additional nonlinear interactions that warrant further exploration. In Fig. 3(c), a frequency of 12.21 MHz appears, corresponding to a period $t_2$ of about 82 ns. At higher microwave power, more pronounced secondary modulations are observed, along with enhanced excitations of the modes $f_4^-$ ($f_4^+$), as shown in the insets of Figs. 3(d) and 3(e). The corresponding FFT spectra (inset) allow us to identify power-dependent soliton periods $t_4$ and $t_2$. As depicted in Fig. 3(f), $t_4$ decreases and eventually saturates with increasing power, while $t_2$ goes through a maximum and then slightly declines at 10 W. We attributed these power-dependent variations in soliton periods to the renormalization of the spin wave dispersion. With increasing power, the precession angle of the saturation magnetization ($M_s$) increases, leading to a downshift of the spin wave frequency and a larger group velocity, which accounts for the increased beating frequency $\Delta f_4$ (i.e., $f_p - f_4^-$) and reduced $t_4$. For the increase in $t_2$, i.e., the decrease of $\Delta f_2$, the frequency spacing between the $n_1$ and $n_1 - 1$ ($n_1 + 1$) branches, we considered that it originates from the frequency compression effect. In the strongly nonlinear regime, particularly at higher powers, nonlinear effects induce frequency modulation, which may cause a reduction in the frequency spacing between adjacent modes. As the microwave power continues to increase, the nonlinear effects finally saturate, and the frequency spacing will not continue to decrease and may even slightly increase, exhibiting a saturation effect.

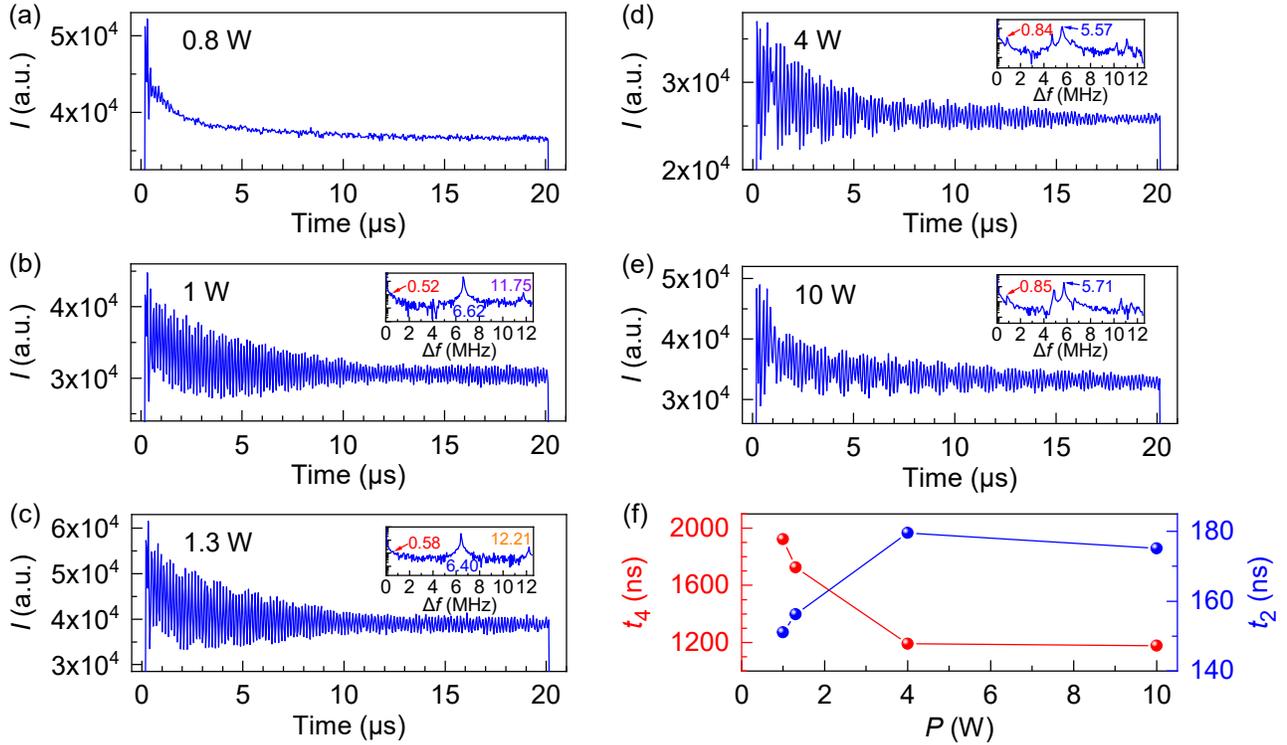

FIG. 3. TR-BLS spectra under different nominal microwave powers. (a) 0.8 W. (b) 1 W. (c) 1.3 W. (d) 4 W. (e) 10 W. The insets show the corresponding FFT results. Measurements were performed at $f_p = 3.6$ GHz, $w = 20$ μs, and $T = 40$ μs. (f) Power-dependent period ($t_1$ and $t_2$) extracted from the FFT results of (b)-(e).

To explore the spatial characteristics and non-reciprocal propagation dynamics of the MSSW solitons, we conducted time- and space-resolved BLS spectroscopy measurements. As illustrated in Fig. 1(a), we scanned the laser spot along both the y- and x-axis to capture the soliton propagation dynamics. The corresponding time-dependent intensity contour plot along the y-axis (the length of YIG), is shown in Fig. 4(a), where distinct soliton trains are clearly observed in the time domain. From these measurements, we can extract the group velocities of the spin waves traveling along the –y- and y-axis, which are 833 m/s and 1240 m/s, respectively, as indicated by the two black arrows. To further analyze the spatial dependence of the normalized Stokes intensity ($I_{norm}$), we examined the profile at $t = 0.567$ μs [marked by the white dotted line in Fig. 4(a)], as shown in Fig. 4(b). The spatial intensity exhibits a superposition of exponential decay and periodic spatial modulation. Fitting the data with the formula $I_{norm} = Ae^{-jy} + B\sin(\Delta ky + \varphi)e^{-my}$ yields a wavevector difference between the two interference magnon modes ($\Delta k_4$) of 0.079 μm$^{-1}$, corresponding to a wavelength $\lambda_4 = 2\pi/\Delta k_4 = 0.08$ mm, as indicated by the dotted blue lines and arrows. The fitting result is shown by the solid red curve. Here, the MSSW soliton can be considered as a quasiparticle, with a wavevector $\Delta k_4$ (0.079 μm$^{-1}$) and frequency $\Delta f_4$ (97.35 MHz) as mentioned in Fig. 2(e), exhibiting periodic modulation in both space and time domains. Additionally, the MSSW soliton trains can be observed in both the Stokes and anti-Stokes BLS signals, which further demonstrates the quasiparticle nature of the soliton. The time- and space-resolved BLS intensity contour plot for the anti-Stokes side, shown in Supplemental Material [51], exhibits behavior similar to that of the Stokes side.

When the laser spot was moved along the $x$-axis (the width of YIG), as shown in Fig. 4(c), soliton patterns were again observed. Unlike the case along the $y$-axis, the internal field distribution is inhomogeneous across the strip, resulting in soliton periods and group velocities that vary with position along the $x$-axis. Moreover, the nonuniform magnetization drives the generation of transverse modes. The wave confinement along the $x$-direction leads to the quantization of $k_x$, given by the expression $k_x = l\pi/w_{\text{eff}}$, where $l$ is an integer (1, 2, 3...) and $w_{\text{eff}}$ represents the effective width that defines the quantization of the spin wave [54, 55]. By analyzing the profile at $t = 0.408$ μs (the white dotted line in Fig. 4(c)), we found that the intensity shows an overall downward trend as we moved the laser away from the microwave input port, as shown in Fig. 4(d). As electromagnetic waves propagate along the length of the microstrip line, losses accumulate in direct proportion to the transmission distance, leading to signal strength decay. Furthermore, we also observed a spatially periodic intensity modulation across the film with $\lambda_x = 0.14$ mm, indicating the presence of the transverse soliton with $\Delta k_x = 2\pi/\lambda_x = 0.045$ μm$^{-1}$. Notably, the spatial intensity distributions along and across the YIG stripe also exhibit periodic time-dependent variations, as shown in Supplemental Material [51].

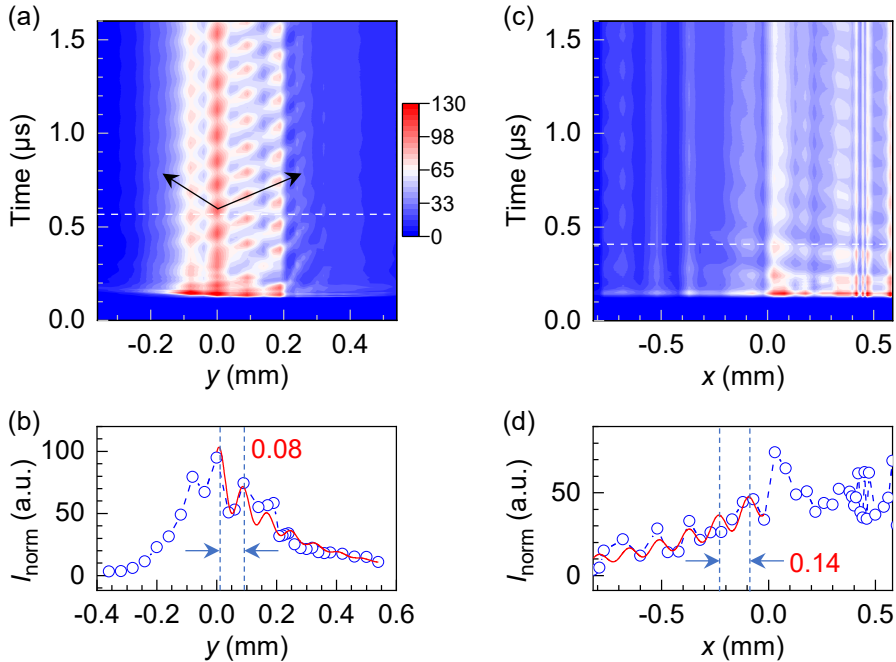

FIG. 4. Time- and space-resolved BLS spectra of the Stokes side. Time-resolved normalized intensity contour plot along the (a) $y$- and (c) $x$-axis. They share the same color bar. The two black arrows denote the propagation of spin waves, from which the group velocity can be extracted. The intensity profile at $t = 0.567$ μs for the (b) $y$- and $t = 0.408$ μs for the (d) $x$-axis, indicated by the white dotted lines in (a) and (c), respectively. The red solid curves in (b) and (d) represent the exponential decay and damped sinc fitting result [$I_{\text{norm}} = Ae^{-jy} + B \sin(\Delta ky + \varphi)e^{-my}$]. The numbers mark the corresponding wavelength ($\lambda = 2\pi/\Delta k$), i.e., the distance between the nearest intensity maxima here, which are indicated by two dotted blue lines and arrows. Measurements were performed at $f_p = 3.6$ GHz, $w = 1.5$ μs, $T = 3$ μs, and $P = 10$ W.

In conclusion, we had resolved the longstanding issue on the existence of MSSW solitons in YIG thin films by advanced spatiotemporally resolved BLS spectroscopy. Our work provides direct experimental evidence of MSSW soliton formation, showing that these solitons exhibit unique long-period trains, with multiple periodicities arising from nonlinear four-magnon scattering processes. These nonlinear interactions lead to mode beating, resulting in an exotic comb-like frequency spectrum that had not been previously observed. We demonstrated that soliton periods can be tuned through microwave power, reflecting the renormalization of spin wave dispersion. Our findings reveal the quasiparticle nature of MSSW solitons, characterized by well-defined wavevectors and frequencies, with periodic modulation emerging in both space and time domains. Importantly, we observed the emergence of transverse solitons across the YIG stripe, manifesting a new form of soliton behavior with spatial periodicity. Micromagnetic simulations agree well with experimental findings. These results not only resolve a critical issue in soliton physics but also deepen our understanding of nonlinear spin-wave dynamics. Our work lays the foundation for the development of advanced nonlinear magnonic devices, enabling the use of multi-periodic MSSW soliton structures in wave-based information processing technologies.

## ACKNOWLEDGMENTS

J. Z. acknowledges the funding support from Research Equipment Development Project of Chinese Academy of Sciences (YJKYYQ20210001), the CAS Project for Young Scientists in Basic Research (YSBR-120), Chinese Academy of Sciences - the Scientific and Technological Research Council of TÜRKİYE Joint Research Projects (172111KYSB20210004), the CAS Interdisciplinary Innovation Team, and National Natural Science Foundation of China (12074371). P. Y. was supported by the National Key R&D Program under Contract No. 2022YFA1402802 and the National Natural Science Foundation of China (Grants No. 12434003, No. 12374103, and No. 12074057).

[1] V.V. Kruglyak, S.O. Demokritov, and D. Grundler, Magnonics, J. Phys. D: Appl. Phys. **43**, 264001 (2010).
[2] B. Lenk, H. Ulrichs, F. Garbs, and M. Münzenberg, The building blocks of magnonics, Phys. Rep. **507**, 107 (2011).
[3] P. Pirro, V.I. Vasyuchka, A.A. Serga, and B. Hillebrands, Advances in coherent magnonics, Nat. Rev. Mater. **6**, 1114 (2021).
[4] B. Flebus, D. Grundler, B. Rana, Y. Otani, I. Barsukov, A. Barman, G. Gubbiotti, P. Landeros, J. Akerman, U. Ebels, P. Pirro, V.E. Demidov, K. Schultheiss, G. Csaba, Q. Wang, F. Ciubotaru, D.E. Nikonov, P. Che, R. Hertel, T. Ono, D. Afanasiev, J. Mentink, T. Rasing, B. Hillebrands, S.V. Kusminskiy, W. Zhang, C.R. Du, A. Finco, T. van der Sar, Y.K. Luo, Y. Shiota, J. Sklenar, T. Yu, and J.W. Rao, The 2024 magnonics roadmap, J. Phys.: Condens. Matter **36**, 363501 (2024).
[5] S.A. Wolf, D.D. Awschalom, R.A. Buhrman, J.M. Daughton, S. von Molnár, M.L. Roukes, A.Y. Chtchelkanova, and D.M. Treger, Spintronics: A spin-based electronics vision for the future, Science **294**, 1488 (2001).
[6] A.V. Chumak, V.I. Vasyuchka, A.A. Serga, and B. Hillebrands, Magnon spintronics, Nat. Phys. **11**, 453 (2015).
[7] A.M. Kosevich, B.A. Ivanov, and A.S. Kovalev, Magnetic Solitons, Phys. Rep. **194**, 117 (1990).
[8] M. Chen, M.A. Tsankov, J.M. Nash, and C.E. Patton, Backward-volume-wave microwave-envelope solitons in yttrium iron garnet films, Phys. Rev. B **49**, 12773 (1994).
[9] M.A. Tsankov, M. Chen, and C.E. Patton, Forward volume wave microwave envelope solitons in yttrium iron garnet films: Propagation, decay, and collision, J. Appl. Phys. **76**, 4274 (1994).
[10] A.B. Ustinov, V.E. Demidov, A.V. Kondrashov, B.A. Kalinikos, and S.O. Demokritov, Observation of the chaotic spin-wave soliton trains in magnetic films, Phys. Rev. Lett. **106**, 017201 (2011).
[11] A.S. Bir, S.V. Grishin, O.I. Moskalenko, A.N. Pavlov, M.O. Zhuravlev, and D.O. Ruiz, Experimental observation of ultrashort hyperchaotic dark multisoliton complexes in a magnonic active ring resonator, Phys. Rev. Lett. **125**, 083903 (2020).
[12] A.S. Bir, S.V. Grishin, A.A. Grachev, O.I. Moskalenko, A.N. Pavlov, D.V. Romanenko, V.N. Skorokhodov, and S.A. Nikitov, Direct electric current control of hyperchaotic packets of dissipative dark envelope solitons in a magnonic crystal active ring resonator, Phys. Rev. Appl. **21**, 044008 (2024).
[13] S.O. Demokritov, A.A. Serga, V.E. Demidov, B. Hillebrands, M.P. Kostylev, and B.A. Kalinikos, Experimental observation of symmetry-breaking nonlinear modes in an active ring, Nature **426**, 159 (2003).
[14] V.E. Demidov, S. Urazhdin, H. Ulrichs, V. Tiberkevich, A. Slavin, D. Baither, G. Schmitz, and S.O. Demokritov, Magnetic nano-oscillator driven by pure spin current, Nat. Mater. **11**, 1028 (2012).
[15] M.B. Jungfleisch, W. Zhang, J. Sklenar, J. Ding, W. Jiang, H. Chang, F.Y. Fradin, J.E. Pearson, J.B. Ketterson, V. Novosad, M. Wu, and A. Hoffmann, Large spin-wave bullet in a ferrimagnetic insulator driven by the spin Hall effect, Phys. Rev. Lett. **116**, 057601 (2016).
[16] A. Houshangh, R. Khymyn, H. Fulara, A. Gangwar, M. Haidar, S.R. Etesami, R. Ferreira, P.P. Freitas, M. Dvornik, R.K. Dumas, and J. Akerman, Spin transfer torque driven higher-order propagating spin waves in nano-contact magnetic tunnel junctions, Nat. Commun. **9**, 4374 (2018).
[17] S.M. Mohseni, S.R. Sani, J. Persson, T.N.A. Nguyen, S. Chung, Y. Pogoryelov, P.K. Muduli, E. Iacocca, A. Eklund, R.K. Dumas, S. Bonetti, A. Deac, M.A. Hoefer, and J. Åkerman, Spin torque-generated magnetic droplet solitons, Science **339**, 1295 (2013).
[18] S. Chung, Q.T. Le, M. Ahlberg, A.A. Awad, M. Weigand, I. Bykova, R. Khymyn, M. Dvornik, H. Mazraati, A. Houshang, S. Jiang, T.N.A. Nguyen, E. Goering, G. Schütz, J. Gräfe, and J. Åkerman, Direct observation of Zhang-Li torque expansion of magnetic droplet solitons, Phys. Rev. Lett. **120**, 217204 (2018).
[19] S. Jiang, S. Chung, M. Ahlberg, A. Frisk, R. Khymyn, Q.T. Le, H. Mazraati, A. Houshang, O. Heinonen, and J. Åkerman, Magnetic droplet soliton pairs, Nat. Commun. **15**, 2118 (2024).
[20] R.K. Kummamuru, and Y.-A. Soh, Electrical effects of spin density wave quantization and magnetic domain walls in chromium, Nature **452**, 859 (2008).

[21] G. Catalan, J. Seidel, R. Ramesh, and J.F. Scott, Domain wall nanoelectronics, Rev. Mod. Phys. **84**, 119 (2012).

[22] A. Wachowiak, J. Wiebe, M. Bode, O. Pietzsch, M. Morgenstern, and R. Wiesendanger, Direct observation of internal spin structure of magnetic vortex cores, Science **298**, 577 (2002).

[23] B. Van Waeyenberge, A. Puzic, H. Stoll, K.W. Chou, T. Tyliszczak, R. Hertel, M. Faehnle, H. Brueckl, K. Rott, G. Reiss, I. Neudecker, D. Weiss, C.H. Back, and G. Schuetz, Magnetic vortex core reversal by excitation with short bursts of an alternating field, Nature **444**, 461 (2006).

[24] X.Z. Yu, Y. Onose, N. Kanazawa, J.H. Park, J.H. Han, Y. Matsui, N. Nagaosa, and Y. Tokura, Real-space observation of a two-dimensional skyrmion crystal, Nature **465**, 901 (2010).

[25] P. Milde, D. Köhler, J. Seidel, L.M. Eng, A. Bauer, A. Chacon, J. Kindervater, S. Mühlbauer, C. Pfleiderer, S. Buhrandt, C. Schütte, and A. Rosch, Unwinding of a skyrmion lattice by magnetic monopoles, Science **340**, 1076 (2013).

[26] A.N. Bogdanov, and C. Panagopoulos, Physical foundations and basic properties of magnetic skyrmions, Nat. Rev. Phys. **2**, 492 (2020).

[27] P.J. Ackerman, and Smalyukh, II, Static three-dimensional topological solitons in fluid chiral ferromagnets and colloids, Nat. Mater. **16**, 426 (2017).

[28] Y. Liu, W. Hou, X. Han, and J. Zang, Three-dimensional dynamics of a magnetic hopfion driven by spin transfer torque, Phys. Rev. Lett. **124**, 127204 (2020).

[29] F. Zheng, N.S. Kiselev, F.N. Rybakov, L. Yang, W. Shi, S. Blugel, and R.E. Dunin-Borkowski, Hopfion rings in a cubic chiral magnet, Nature **623**, 718 (2023).

[30] G. Finocchio, M. Ricci, R. Tomasello, A. Giordano, M. Lanuzza, V. Puliafito, P. Burrascano, B. Azzerboni, and M. Carpentieri, Skyrmion based microwave detectors and harvesting, Appl. Phys. Lett. **107**, 262401 (2015).

[31] R. Sharma, R. Mishra, T. Ngo, Y.-X. Guo, S. Fukami, H. Sato, H. Ohno, and H. Yang, Electrically connected spin-torque oscillators array for 2.4GHz WiFi band transmission and energy harvesting, Nat.. Commun. **12**, 2924 (2021).

[32] X. Zhang, M. Ezawa, and Y. Zhou, Magnetic skyrmion logic gates: conversion, duplication and merging of skyrmions, Sci. Rep. **5**, 9400 (2015).

[33] Z. Luo, A. Hrabec, T.P. Dao, G. Sala, S. Finizio, J. Feng, S. Mayr, J. Raabe, P. Gambardella, and L.J. Heyderman, Current-driven magnetic domain-wall logicl, Nature **579**, 214 (2020).

[34] S.S.P. Parkin, M. Hayashi, and L. Thomas, Magnetic domain-wall racetrack memory, Science **320**, 190 (2008).

[35] P. Fedorov, I. Soldatov, V. Neu, R. Schaefer, O.G. Schmidt, and D. Karnaushenko, Self-assembly of Co/Pt stripes with current-induced domain wall motion towards 3D racetrack devices, Nat. Commun. **15**, 2048 (2024).

[36] R. Lavrijsen, J.H. Lee, A. Fernández-Pacheco, D. Petit, R. Mansell, and R.P. Cowburn, Magnetic ratchet for three-dimensional spintronic memory and logic, Nature **493**, 647 (2013).

[37] M. Romera, P. Talatchian, S. Tsunegi, F.A. Araujo, V. Cros, P. Bortolotti, J. Trastoy, K. Yakushiji, A. Fukushima, H. Kubota, S. Yuasa, M. Ernoult, D. Vodenicarevic, T. Hirtzlin, N. Locatelli, D. Querlioz, and J. Grollier, Vowel recognition with four coupled spin-torque nano-oscillators, Nature **563**, 230 (2018).

[38] G. Marcucci, D. Pierangeli, and C. Conti, Theory of neuromorphic computing by waves: Machine learning by rogue waves, dispersive shocks, and solitons, Phys. Rev. Lett. **125**, 093901 (2020).

[39] M. Zahedinejad, A.A. Awad, S. Muralidhar, R. Khymyn, H. Fulara, H. Mazraati, M. Dvornik, and J. Akerman, Two-dimensional mutually synchronized spin Hall nano-oscillator arrays for neuromorphic computing, Nat. Nanotechnol. **15**, 47 (2020).

[40] J.L. Chen, H.M. Yu, and G. Gubbiotti, Unidirectional spin-wave propagation and devices, J. Phys. D: Appl. Phys. **55**, 123001 (2022).

[41] M.J. Hurben, and C.E. Patton, Theory of magnetostatic waves for in-plane magnetized isotropic films, J. Magn. Magn. Mater. **139**, 263 (1995).

[42] M.J. Lighthill, Contributions to the theory of waves in non-linear dispersive systems, J. Inst. Math. Its Appl. **1**, 269 (1965).

[43] J.M. Nash, P. Kabos, R. Staudinger, and C.E. Patton, Phase profiles of microwave magnetic envelope solitons, J. Appl. Phys. **83**, 2689 (1998).

[44] B.A. Kalinikos, N.G. Kovshikov, P.A. Kolodin, and A.N. Slavin, Observation of dipole-exchange spin wave solitons in tangentially magnetised ferromagnetic films, Solid State Commun. **74**, 989 (1990).

[45] T. Iwata, T. Eguchi, and K. Sekiguchi, Generation of spin-wave soliton using magnetostatic surface mode, IEEE International Magnetic Conference - Short Papers, Sendai, Japan, 2023, pp. 1.

[46] M. Madami, G. Gubbiotti, S. Tacchi, and G. Carlotti, Chapter two - Application of microfocused Brillouin light scattering to the study of spin waves in low-dimensional magnetic systems, in Solid State Phys., Academic Press, 2012, pp. 79.

[47] T. Sebastian, K. Schultheiss, B. Obry, B. Hillebrands, and H. Schultheiss, Micro-focused Brillouin light scattering: imaging spin waves at the nanoscale, Front. Phys. **3**, 35 (2015).

[48] F. Kargar, and A.A. Balandin, Advances in Brillouin-Mandelstam light-scattering spectroscopy, Nat. Photonics **15**, 720 (2021).

[49] B.A. Kalinikos, and A.N. Slavin, Theory of dipole-exchange spin wave spectrum for ferromagnetic films with mixed exchange boundary conditions, J. Solid State Phys. **19**, 7013 (1986).
[50] A.N. Slavin, and I.V. Rojdestvenski, "Bright" and "dark" spin wave envelope solitons in magnetic films, IEEE Trans. Magn. **30**, 37 (1994).
[51] See Supplemental Material [url] for different spatial intensity profiles as a function of time and the time- and space-resolved BLS intensity of the anti-Stokes side, together with the estimation of the microwave power delivered to the sample, which includes Ref. [52].
[52] A. Vansteenkiste, J. Leliaert, M. Dvornik, M. Helsen, F. Garcia-Sanchez, and B. Van Waeyenberge, The design and verification of MuMax3, AIP Adv. **4**, 107133 (2014).
[53] M. Remoissenet, Waves called solitons: Concepts and experiments, Springer, Berlin, 1995.
[54] C. Bayer, J.P. Park, H. Wang, M. Yan, C.E. Campbell, and P.A. Crowell, Spin waves in an inhomogeneously magnetized stripe, Phys. Rev. B **69**, 134401 (2004).
[55] P. Pirro, T. Brächer, K. Vogt, B. Obry, H. Schultheiss, B. Leven, and B. Hillebrands, Interference of coherent spin waves in micron-sized ferromagnetic waveguides, Phys. Status Solidi B **248**, 2404 (2011).

# Supplemental Material for "Observation of Magnetostatic Surface Spin Wave Solitons in Yttrium Iron Garnet Thin Film"

Simin Pang,[1,#] Zhengyi Li,[2,4,#] Ziyu Wang,[5,#] Yanpei Lv,[1,3] Feilong Song,[1] Peng Yan,[4,*] and Jun Zhang[1,3,†]

[1]*State Key Laboratory of Semiconductor Physics and Chip Technologies, Institute of Semiconductors, Chinese Academy of Sciences, Beijing 100083, China*
[2]*Jiangsu Key Laboratory of Opto-Electronic Technology, School of Physics and Technology, Nanjing Normal University, Nanjing 210046, China*
[3]*Center of Materials Science and Optoelectronics Engineering, University of Chinese Academy of Sciences, Beijing 100049, China*
[4]*School of Physics and State Key Laboratory of Electronic Thin Films and Integrated Devices, University of Electronic Science and Technology of China, Chengdu 610054, China*
[5]*School of Physics and Laboratory of Zhongyuan Light, Zhengzhou University, Zhengzhou 450001, China*
[#]These authors contributed equally.
**Corresponding Author**
E-mail: [†]zhangjwill@semi.ac.cn; [*]yan@uestc.edu.cn

## 1. Simulated spin wave dispersion and temporal evolution of the spatial distribution of soliton amplitude.

To investigate the relationships between the existing dipole gaps and the observation of MSSW solitons, we first employed Mumax[3] for micromagnetic simulation [1], calculating the spin wave dispersion. The essence of micromagnetic simulation involves numerically solving the Landau-Lifshitz-Gilbert (LLG) equation. The simulated structure is a YIG strip, with dimensions of $400 \times 1 \times 0.4$ μm³, discretized into a grid of $8000 \times 1 \times 8$ cells. The magnetic material parameters are given as follows: exchange stiffness coefficient $A_{ex}$ = 3.5×10$^{-12}$ J/m, saturation magnetization $M_s$ = 1758 Oe, and damping coefficient $\alpha$ = 5×10$^{-4}$. Gradient damping at both ends of the YIG strip serves as absorptive boundary conditions. The external magnetic field of 1000 Oe is applied along the *x*-direction, saturating the magnetization. To effectively excite MSSW, a localized sinc function perturbation field with a cutoff frequency of 25 GHz is applied at the structure's center.

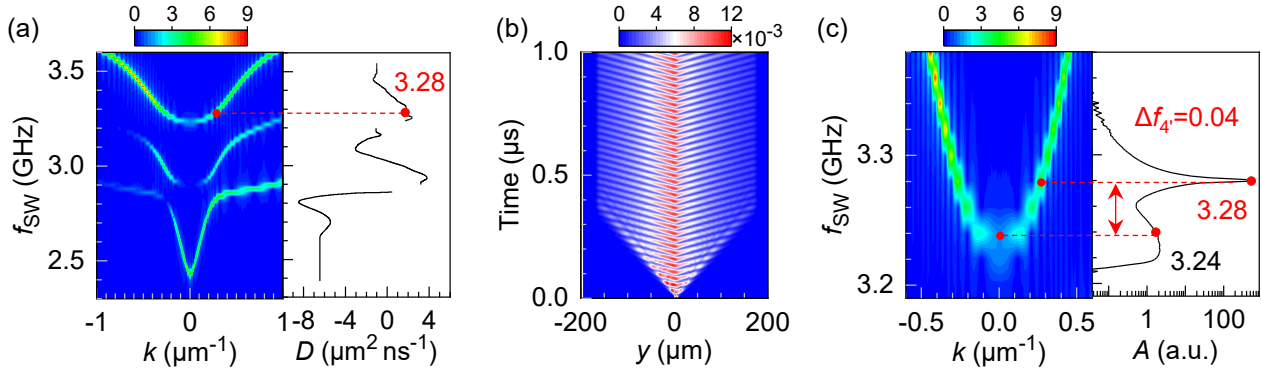

FIG. S1. Simulated temporal evolution of the spatial distribution of spin wave amplitude under continuous excitation at $f_p$ = 3.28 GHz for MSSW soliton. (a) Left panel: color plot of simulated spin wave dispersion relation. Right panel: extracted dispersion parameter $D$ from the left panel. Red dots indicate the pumping frequency of 3.28 GHz. (b) The full temporal evolution of the spatial distribution of spin wave amplitude. (c) Left panel: the zoom-in dispersion extracted from (a). Right panel: the FFT result for the $m_y$ at $y$ = 100 μm. Red dots indicate the pumping frequency $f_p$ = 3.28 GHz and $f_{4'}^-$ = 3.24 GHz. The frequency difference $\Delta f_{4'}$ = 0.04 GHz.

The calculated dispersion is shown in the left panel of Fig. S1(a), exhibiting pronounced dipole gaps, from which we can extract the dispersion parameter $D$ [right panel of Fig. S1(a)]. $f_p$ = 3.28 GHz (red dots) satisfying the Lighthill criterion was chosen for a continuous pump duration of 1000 ns. Figure S1(b) displays the temporal evolution of the spatial distribution of spin wave amplitude ($|m_y|^2 + |m_z|^2$), revealing a distinct envelope that gradually smoothens as time progresses. For the $m_y$ at $y$ = 100 μm, the fast Fourier transform (FFT) result is shown in Fig. S1(c), from which we can obtain two obvious frequencies: the pumping frequency ($f_p$ = 3.28 GHz) and the frequency at the bottom of the upper branch of the dipolar bandgap ($f_{4'}^-$ = 3.24 GHz). The difference frequency between $f_p$ and $f_{4'}^-$ ($\Delta f_{4'}$ = 0.04 GHz) is exactly the reciprocal of the soliton period ($t_{4'} \approx 25$ ns).

Similarly, we performed the simulation using the YIG stripe with an increased thickness of 1.2 μm, discretizing the structure into a grid of 2000 × 1 × 24 cells and considering an anisotropy field of 1300 Oe. The external magnetic field of 680 Oe is applied along the $x$-direction, saturating the magnetization. The red dots in Fig. S2(a) denote the pumping frequency of 3.76 GHz. The FFT result for the $m_y$ at $y$ = 50 μm [right panel in Fig. S2(c)] presents the spectrum with three frequencies from high to low being the frequency $f_{2'}^+$ = 3.84 GHz at the upper branch of the bandgap, the pump frequency $f_p$ = 3.76 GHz, and the lower branch of the bandgap $f_{2'}^-$ = 3.68 GHz, which possess the same wavevector. The frequency difference between $f_p$ and $f_{2'}^+$ ($f_{2'}^-$) is the reciprocal of the soliton period ($t_{2'} \approx 11$ ns).

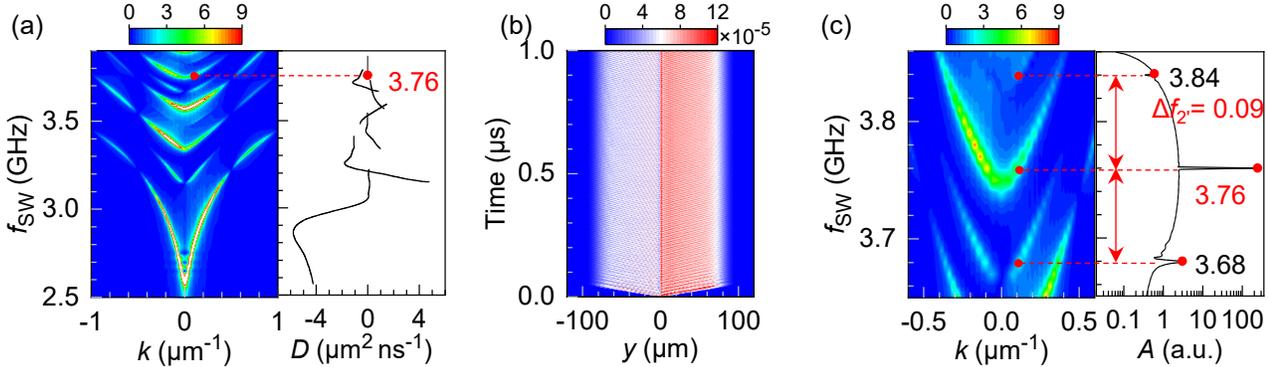

FIG. S2. Simulated temporal evolution of the spatial distribution of spin wave amplitude under continuous excitation at $f_p$ = 3.76 GHz for MSSW soliton. (a) Left panel: color plot of simulated spin wave dispersion relation. Right panel: extracted dispersion parameter $D$ from the left panel. Red dots indicate the pumping frequency of 3.76 GHz. (b) The full temporal evolution of the spatial distribution of spin wave amplitude. (c) Left panel: the zoom-in dispersion extracted from (a). Right panel: the FFT result for the $m_y$ at $y$ = 50 μm. Red dots indicate the pumping frequency $f_p$ = 3.76 GHz, $f_{2'}^+$ = 3.84 GHz, and $f_{2'}^-$ = 3.68 GHz. The frequency difference $\Delta f_{2'}$ = 0.09 GHz.

## 2. Estimation of the microwave power delivered to the sample.

The microwave from the microwave source with a nominal output power of $P_{norminal}$ = 26 dBm (398.1 mW) was directly transferred to the spectrum analyzer. The output power ($P_{in}$) is 23.87 dBm (243.8 mW), with a loss of $P_{loss}$ = 154.3 mW, as shown by the blue curve in Fig. S3. When the $P_{norminal}$ passes through the microstrip resonator we used in the experiment

[Fig. 1(a)] and then to the spectrum analyzer, the output power ($P_{out}$) is -12.46 dBm (0.06 mW), as shown by the red curve in Fig. S3. Considering that the microwave power decays exponentially along the length of the resonator, we can use the formula $P_{out} = P_{in}e^{-\beta d}$ to characterize the power variation, where $\beta$ is attenuation coefficient and $d$ is distance from the input port of the microstrip resonator. When $d = 0$, $P_{out} = P_{in} = 243.8$ mW, while $d = 2$ cm (output port of the microstrip resonator), $P_{out} = 0.06$ mW, from which we can obtain $\beta = 4.15$. Therefore, for the $P_{norminal} = 40$ dBm (10 W) we used in Fig. 2, $P_{in} = P_{norminal} - P_{loss} = 9845.7$ mW. The distance from the input port to the sample position $d = 1.3$ cm, we can thus estimate that the power delivered to the sample is approximately 45 mW. Similarly, for the nominal power used in Fig. 3, we can estimate the applied power to be 2.9 mW, 3.8 mW, 5 mW, and 17.4 mW, for the nominal power 0.8 W, 1 W, 1.3 W, and 4 W used in Fig. 3, respectively.

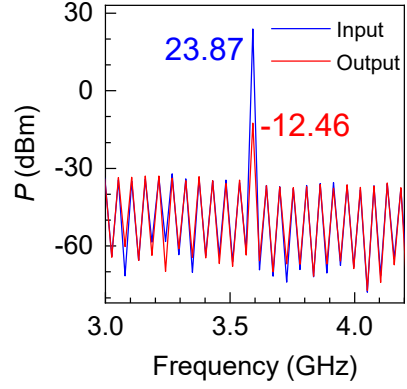

FIG. S3. Spectra obtained by the spectrum analyzer. The pumping frequency $f_p = 3.6$ GHz, and the nominal power is 26 dBm (398.1 mW). The blue curve corresponds to the spectrum with the microwave directly from the microwave source to the spectrum analyzer, while the red one represents the spectrum with the microwave from the microwave source passing through the microstrip resonator and then to the spectrum analyzer.

## 3. Time- and space-resolved BLS intensity contour plot for the anti-Stokes side

Figure S4 shows the time- and space-resolved BLS intensity contour plot for the anti-Stokes side, which behaves similarly to the Stokes one (Fig. 4), with the same group velocity and wavelength for the MSSW soliton and transverse soliton.

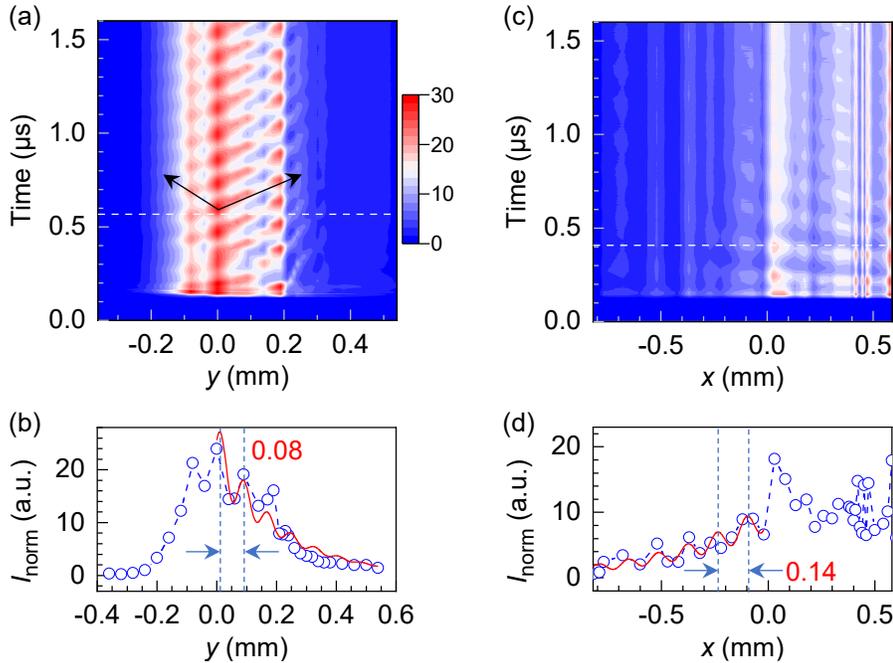

FIG. S4. Time- and space-resolved BLS spectra of the anti-Stokes side. Time-resolved normalized intensity contour plot along the (a) $y$- and (c) $x$-axis. The two black arrows denote the propagation of spin waves, from which the group velocity can be extracted. The intensity profile at $t = 0.567$ μs for the (b) $y$- and $t = 0.408$ μs for the (d) $x$-axis, indicated by the white dotted lines in (a) and (c), respectively. The red solid curves in (b) and (d) represent the exponential decay and damped sinc fitting result [$I_{norm} = Ae^{-jy} + B \sin(\Delta ky + \varphi)e^{-my}$]. The numbers mark the corresponding wavelength ($\lambda = 2\pi/\Delta k$), i.e., the distance between the nearest intensity maxima here, which are indicated by two dotted blue lines and arrows.

## 4. Experimental temporal evolution of the normalized intensity of the Stokes signals along the *y*- and *x*-axis.

As shown in Fig. S5, apart from the periodic modulation in the space domain, the spatial intensity distributions along (*y*-axis) and across (*x*-axis) the YIG stripe also exhibit periodic time-dependent variations.

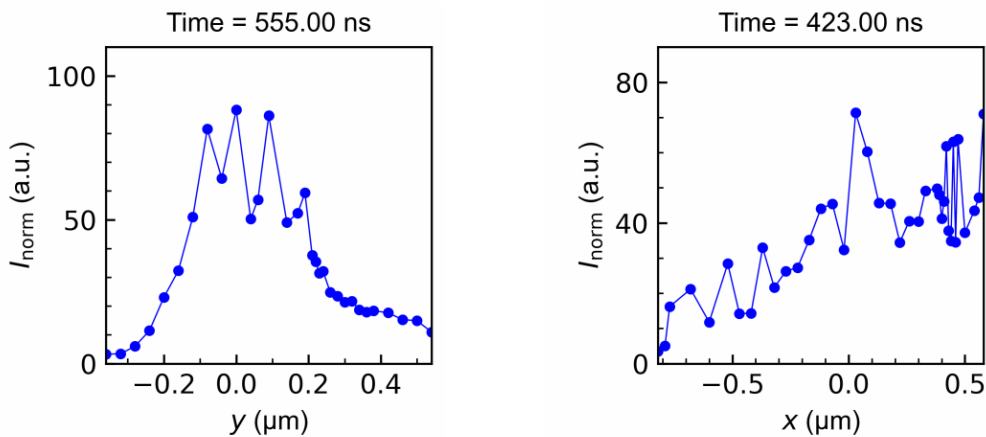

FIG. S5. Experimental temporal evolution of the normalized intensity of the Stokes signals along the *y*- and *x*-axis, extracted from Fig. 4(a) and 4(b), respectively.

---

[1] A. Vansteenkiste, J. Leliaert, M. Dvornik, M. Helsen, F. Garcia-Sanchez, and B. Van Waeyenberge, The design and verification of MuMax3, AIP Adv. **4**, 107133 (2014).